\documentclass[conference]{IEEEtran}
\IEEEoverridecommandlockouts

\usepackage{booktabs}
\usepackage{cite}
\usepackage{amsmath,amssymb,amsfonts}
\usepackage{algorithmic}
\usepackage{graphicx}
\usepackage{textcomp}
\usepackage{xcolor}
\usepackage{multirow}
\IEEEoverridecommandlockouts

\def\BibTeX{{\rm B\kern-.05em{\sc i\kern-.025em b}\kern-.08em
    T\kern-.1667em\lower.7ex\hbox{E}\kern-.125emX}}
\begin{document}

\title{From Elastic to Viscoelastic: An EEMD-Enhanced Pulse Transit Time Model for Robust Blood Pressure Estimation\\
{\footnotesize}
}

\author{
    \IEEEauthorblockN{Boyuan Gu\textsuperscript{1,2}, Yijin Yang\textsuperscript{1}, Shuaiqi Cheng\textsuperscript{1}, and Xiaorong Ding\textsuperscript{1,*}}
    \IEEEauthorblockA{\textit{University of Electronic Science and Technology of China (UESTC)}\\
    }

    \thanks{This work is the pre-print accepted by EMBC 2026. © 2026 IEEE. Personal use of this material is permitted. Permission from IEEE must be obtained for all other uses, in any current or future media, including reprinting/republishing this material for advertising or promotional purposes, creating new collective works, for resale or redistribution to servers or lists, or reuse of any copyrighted component of this work in other works.}
    \thanks{Boyuan Gu (guboyuan79@gmail.com) is with $^1$ UESTC \& $^2$ Tsinghua University SIGS. Yijing Yang, Shuaiqi and Xiaorong Ding are with $^1$ UESTC. $^*$ Xiaorong Ding (xiaorong.ding@uestc.edu.cn) is the corresponding author. }
}

\maketitle

\begin{abstract}
Cuffless blood pressure (BP) estimation based on Pulse Transit Time (PTT) has emerged as a promising solution for continuous health monitoring. However, conventional models relying on the Moens-Korteweg equation often fail during rapid hemodynamic fluctuations, as they assume arterial walls are purely elastic and neglect inherent viscoelasticity. To address this limitation, we propose a physics-informed framework introducing a viscoelastic compensation mechanism. First, raw photoplethysmogram (PPG) signals undergo high-fidelity reconstruction using Modified Akima (Makima) interpolation. Second, a robust Intersecting Tangent Method is applied for precise pulse foot localization. Crucially, we utilize Ensemble Empirical Mode Decomposition (EEMD) to isolate high-frequency Intrinsic Mode Functions (IMFs), defining a ``Viscoelastic Velocity Metric'' to quantify the vascular damping effect ($\eta \cdot \dot{\epsilon}$) typically ignored by elastic models. The framework was rigorously validated on a challenging subset of the MIMIC-II database (364 subjects, 28,525 cardiac cycles) characterized by a high prevalence of hypertension (23.4\%). Experimental results demonstrate medical-grade accuracy, yielding a Root Mean Square Error (RMSE) of 5.22 mmHg for Systolic and 3.65 mmHg for Diastolic BP, with Pearson correlation coefficients ($R > 0.97$). These findings confirm that incorporating viscoelastic features significantly enhances robustness against vascular hysteresis.
\end{abstract}


\section{Introduction}

Continuous, cuffless blood pressure (BP) monitoring is increasingly recognized as a pivotal component of modern healthcare, promising improved hypertension management and cardiovascular risk reduction \cite{b1}. Ubiquitous monitoring of BP via wearable devices offers a solution by allowing unobtrusive, longitudinal tracking, thus detecting hypertensive patterns that escape clinical snapshots \cite{b2}. Recent advances in flexible sensors and miniaturized electronics have spurred the development of portable BP monitors that operate without a cuff \cite{b3} \cite{c1}. However, translating pulse waveform signals into accurate pressure estimates in a calibration-free manner remains a significant engineering challenge.

Pulse Transit Time (PTT) has emerged as the dominant surrogate marker in cuffless BP research \cite{b1,b4}. Based on the physics of wave propagation, PTT is inversely related to arterial stiffness and thus to BP \cite{b5}. Traditional models, derived from the Moens--Korteweg equation, simplify the artery as an ideal \textit{elastic tube} \cite{b6}. This assumption implies a static, one-to-one mapping between PTT and BP. While effective under stable conditions, purely elastic models often struggle to generalize in ambulatory settings, neglecting confounding factors like vascular tone and blood flow dynamics, which leads to significant calibration drift \cite{b8}.

A fundamental limitation of the elastic assumption is that biological arteries are \textit{inherently viscoelastic}. Arterial walls exhibit frequency-dependent behavior: they not only store energy elastically but also dissipate energy via viscosity, creating hysteresis in the pressure--diameter relationship \cite{b9}. Empirical studies confirm that for the same BP level, PTT varies depending on the direction of hemodynamic change (e.g., exercise vs. recovery) \cite{b10}. This ``vascular lag,'' caused by smooth muscle damping, means that standard PTT models systematically underestimate BP during rapid rising phases \cite{b11}. However, explicitly modeling this viscosity (e.g., via differential equations) requires physiological parameters often unavailable in wearable settings. While recent data-driven frameworks have shown promise in extracting complex mappings directly from time-frequency transformed PPG signals using deep learning \cite{c2}, a robust surrogate capable of explicitly quantifying this viscous damping from the signal itself in a physics-informed manner is urgently needed.

\begin{figure*}[t]
    \centering
    \includegraphics[width=\linewidth]{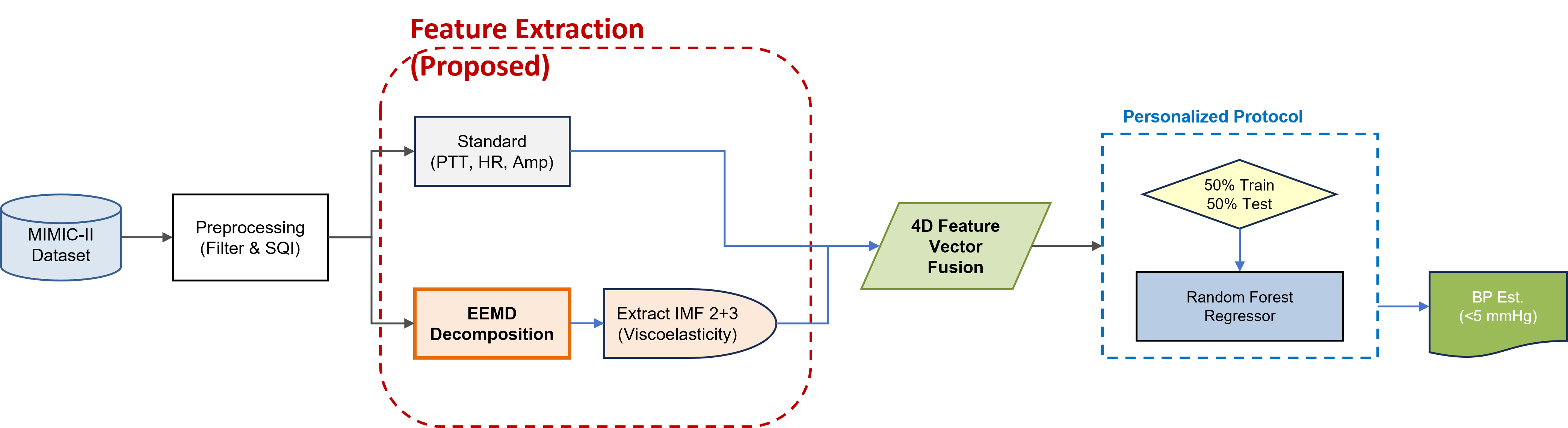}
    \caption{\textbf{System Overview.} The proposed framework integrates an EEMD-based decomposition branch (orange box). By extracting high-frequency Intrinsic Mode Functions (IMF 2-3) as viscoelastic compensation terms, the model fuses a 4D feature vector to drive a robust Random Forest regressor.}
    \label{fig:framework}
    \vspace{-3mm}
\end{figure*}

Bridging this physiological gap requires advanced signal decomposition. Since vascular viscosity is proportional to the arterial strain rate ($\eta \cdot d\epsilon/dt$), its effects are encoded in the high-frequency morphological details of the photoplethysmogram (PPG), particularly in the rapid systolic upstroke. Standard linear filtering often smooths out these subtle viscous markers. To address this, Ensemble Empirical Mode Decomposition (EEMD) offers a powerful adaptive tool \cite{b12}. By decomposing the PPG into IMFs, EEMD can isolate specific frequency bands associated with vascular wall kinetics, separating them from low-frequency respiratory baselines and high-frequency noise \cite{b13}, thereby preserving the transient features essential for viscoelastic modeling.

In this work, we propose a physics-informed viscoelastic PTT framework.  Our specific contributions are: (1) We introduce a \textbf{Viscoelastic Velocity Metric} derived from EEMD IMFs to quantify the vascular damping effect ($\eta \cdot \dot{\epsilon}$); (2) We propose a robust \textbf{Intersecting Tangent Method} for pulse foot localization; and (3) We validate the framework on a \textbf{clinically challenging subset} of the MIMIC-II database (23.4\% hypertensive), demonstrating significantly enhanced robustness against rapid hemodynamic fluctuations.

\section{Methodology}
\label{sec:methodology}

The overall architecture (Fig. \ref{fig:framework}) processes synchronized ECG and PPG data through three stages:

(1) \textbf{Preprocessing and Reconstruction}, where raw signals undergo filtering and high-fidelity up-sampling; 
(2) \textbf{Hybrid Feature Extraction}, the core innovation fusing standard hemodynamic parameters (PTT, HR, Amp) with novel EEMD-derived viscoelastic features; and 
(3) \textbf{Personalized Modeling}, utilizing a Random Forest regressor. 
This hybrid design specifically aims to compensate for the vascular viscoelastic hysteresis typically neglected by elastic models.

\subsection{High-Fidelity Signal Reconstruction}
\label{subsec:reconstruction}
The precision of PTT-based estimation is fundamentally limited by the temporal resolution of the PPG signal. Standard clinical datasets often provide recordings at sampling rates (e.g., 125 Hz) insufficient for capturing millisecond-level transit time variations \cite{b14}.
To mitigate quantization errors, we employ \textit{Modified Akima (Makima) interpolation} to up-sample the raw signal by a factor of 10 (to 1250 Hz) \cite{b15}. Unlike cubic splines, which suffer from the Runge phenomenon (overshoots) near steep gradients, Makima constructs a piecewise polynomial based on weighted local slopes, preserving the true morphology of the systolic upstroke.

\subsection{Robust Feature Localization}
\label{subsec:localization}
\begin{figure}
    \centering
    \includegraphics[width=0.85\linewidth]{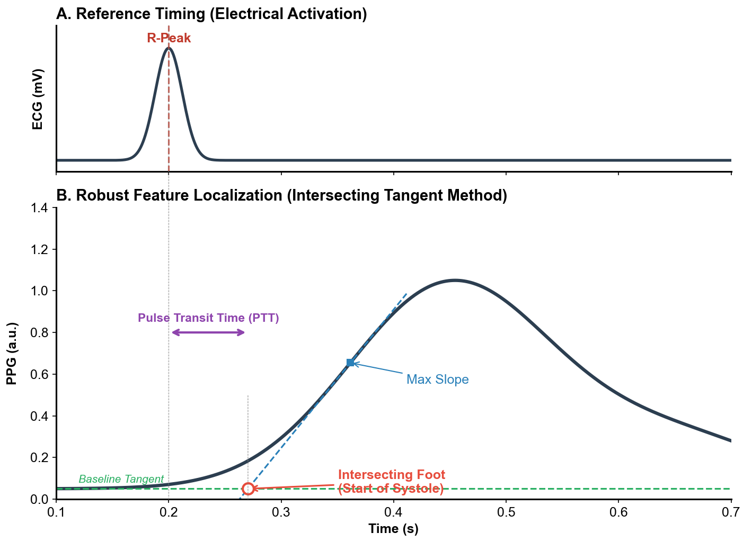}
    \vspace{-2mm}
    \caption{\textbf{Robust Feature Localization.} The Intersecting Tangent Method locates the pulse foot by intersecting the maximum upstroke gradient with the diastolic baseline, ensuring hemodynamic stability.}
    \label{fig1PTT}
\end{figure}

Accurate PTT calculation relies on identifying the precise time delay between the heart activation and the mechanical arrival of the pulse at the periphery.
We define the Start Point as the R-peak of the ECG and the End Point as the PPG signal. Conventional methods often rely on systolic peaks, which are susceptible to wave reflections and motion artifacts. To address this, We propose a robust \textit{Intersecting Tangent Method} (Fig. \ref{fig1PTT}) to localize the PPG foot by: (1) \textbf{Slope Identification:} locating the peak upstroke velocity ($P'_{max}$); and (2) \textbf{Intersection:} crossing the tangent at $P'_{max}$ with the diastolic minimum level. The resulting fiducial point is inherently resistant to baseline drift and noise.

\subsection{Physics-Informed Viscoelastic Compensation}
\label{subsec:eemd_model}
Conventional PTT models assume arterial walls are purely elastic. However, biological tissues are viscoelastic, modeled here by the Kelvin-Voigt constitutive equation:
\begin{equation}
    \sigma(t) = E \cdot \epsilon(t) + \eta \cdot \frac{d\epsilon(t)}{dt}
    \label{eq:kelvin_voigt}
\end{equation}
where $\sigma(t)$ is arterial pressure, $E$ is elastic modulus, and $\eta$ is viscosity. Standard models neglect the viscous damping term ($\eta \cdot \dot{\epsilon}$), causing estimation hysteresis \cite{b9}.

Physiologically, viscous dissipation ($\propto \dot{\epsilon}$) peaks during the rapid systolic upstroke, where the arterial wall deformation rate is maximal. 

Spectral analysis reveals that these transient dynamics concentrate energy in the 2nd--5th harmonics (approx. 5--15 Hz), distinct from the elastic fundamental wave ($<2$ Hz). 

Therefore, we specificially isolate $IMF_2$ and $IMF_3$ from the EEMD decomposition to capture these \textit{Viscoelastic Components}, effectively segregating them from low-frequency elastic dominance and high-frequency sensor noise.

To quantify damping intensity, we introduce the \textbf{Viscoelastic Velocity Metric ($V_{visco}$)}, representing the energy of the rapid deformation rate:
\begin{equation}
    V_{visco} = \ln \left( \frac{1}{K} \sum_{k=1}^{K} \left( \Delta [IMF_2(k) + IMF_3(k)] \right)^2 \right)
    \label{eq:v_visco}
\end{equation}
where $\Delta$ denotes the discrete difference operator. This metric serves as a proxy for the hysteresis loop area. 

Finally, we construct a 4-dimensional feature vector $\mathbf{x} = [1/PTT, V_{visco}, \text{HR}, \text{Amp}]$ to train a \textbf{Random Forest} regressor. We configured the model with 100 decision trees and a maximum depth of 15 to minimize the Mean Squared Error (MSE) during calibration.

\begin{figure}[t]
    \centering
    \includegraphics[width=\linewidth]{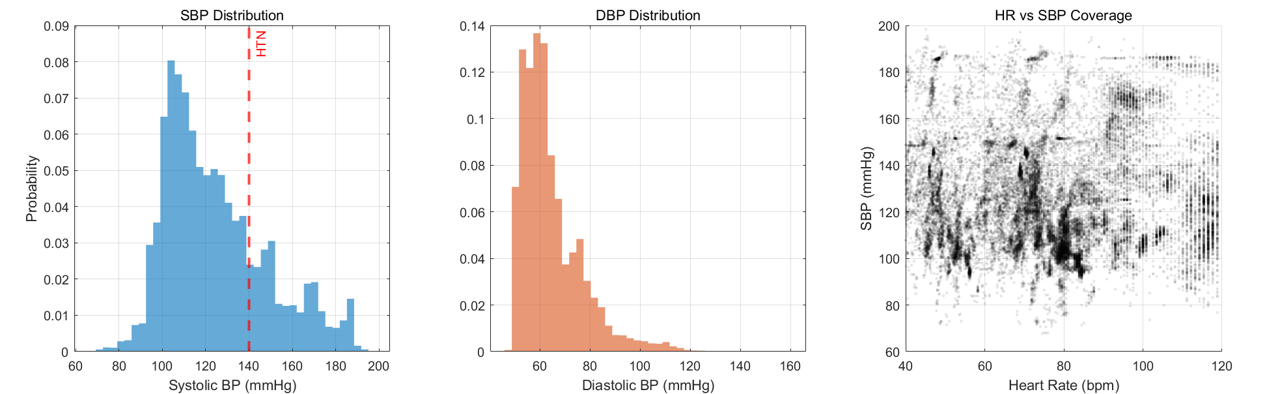} 
    \caption{\textbf{Statistical Distribution.} (Left \& Middle) Histograms show a wide BP range including hypertensive cases ($>140$ mmHg). (Right) HR-SBP scatter plot confirms data diversity and decoupling.}
    \label{fig:data_dist}
\end{figure}

\section{Results and Discussion}
\label{sec:results}

\subsection{Experimental Setup and Dataset Protocol}
\label{subsec:dataset}

To rigorously evaluate the physical validity of the proposed viscoelastic compensation mechanism, we curated a high-fidelity dataset from the MIMIC-II Clinical Database. Unlike standard datasets dominated by normotensive subjects, this study specifically targeted a population with high clinical complexity to challenge the algorithm's robustness beyond the linear elastic regime.

The curated subset comprises \textbf{364 unique subjects}, totaling \textbf{28,525 valid cardiac cycles}. The data selection criteria prioritized signal quality and hemodynamic diversity. The dataset covers an exceptionally wide dynamic range, with Systolic Blood Pressure (SBP) spanning from 68 to 198 mmHg and Diastolic Blood Pressure (DBP) from 46 to 160 mmHg.


Crucially, the dataset exhibits a challenging distribution: while 49.3\% of the samples fall within the normal range, a significant portion (\textbf{23.4\%}) are classified as Hypertensive (SBP $>140$ mmHg). This high prevalence of hypertension is instrumental in verifying our hypothesis: specifically, whether the EEMD-extracted viscous features can effectively correct the PTT model when vascular stiffness increases and the purely elastic assumption typically fails.

For evaluation, we adopted a \textbf{personalized calibration protocol} with a chronological 50\% training and 50\% testing split. 
This chronological 50\% split ensures the training set covers the wide hemodynamic state space inherent in ICU patients, allowing the regressor to efficiently learn non-linear hysteresis patterns across the full dynamic range.
The performance was assessed using Root Mean Square Error (RMSE) and Pearson Correlation Coefficient ($R$), strictly adhering to the IEEE Standard 1708 and AAMI standards.

\subsection{Overall Estimation Accuracy}
\label{subsec:accuracy}

\begin{figure}[t]
    \centering
    \includegraphics[width=\linewidth]{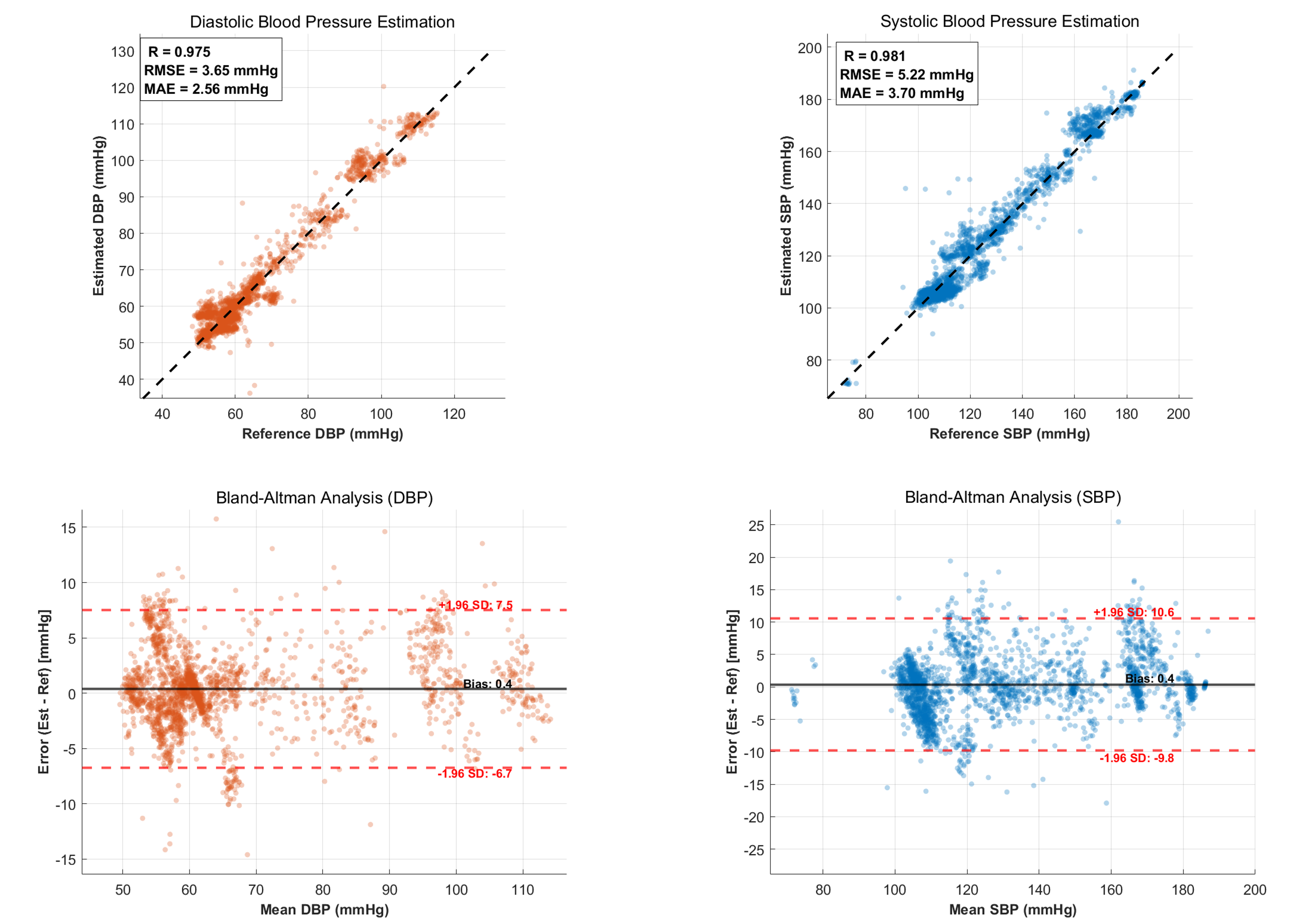} 
    \caption{\textbf{Overall Performance.} (Top) Regression against reference. (Bottom) Bland-Altman plots demonstrating robustness across the wide/hypertensive range.}
    \label{fig:results}
\end{figure}

The overall estimation performance of the proposed model on the testing set is visualized in Fig. \ref{fig:results}.

First, the regression analysis reveals a strong linear correlation between the estimated BP and the reference standard. The model achieved a Pearson correlation coefficient ($R$) of \textbf{0.981} for SBP and \textbf{0.975} for DBP. Quantitatively, the prediction errors were minimized, yielding a Root Mean Square Error (RMSE) of \textbf{5.22 mmHg} for SBP and \textbf{3.65 mmHg} for DBP. These metrics indicate that the EEMD-derived viscoelastic features successfully capture the underlying hemodynamic variations, preventing the saturation effects typically observed in traditional PTT models at high pressures.

Second, the agreement between the estimated and reference values was further assessed using Bland-Altman analysis (Fig. \ref{fig:results}, Bottom Row). The results demonstrate a negligible systematic bias of \textbf{0.4 mmHg} for both SBP and DBP, suggesting no significant over- or under-estimation tendency. The 95\% Limits of Agreement (LoA) were calculated as $[-9.8, 10.6]$ mmHg for SBP and $[-6.7, 7.5]$ mmHg for DBP. Crucially, these results satisfy the AAMI criterion (ME $\le 5$ mmHg, SD $\le 8$ mmHg), confirming the model's utility for medical-grade wearable monitoring.

It is worth noting that strictly enforcing a \textit{chronological} split (rather than random shuffling) imposes a rigorous test on generalization. 
Since MIMIC-II subjects often undergo rapid physiological drifts due to interventions, the testing set frequently presents hemodynamic states unseen in the training phase. 
The high accuracy maintained under this protocol confirms that the EEMD features capture intrinsic vascular properties rather than merely memorizing local waveform morphology.

\begin{figure}[t]
    \centering
    \includegraphics[width=0.9\linewidth]{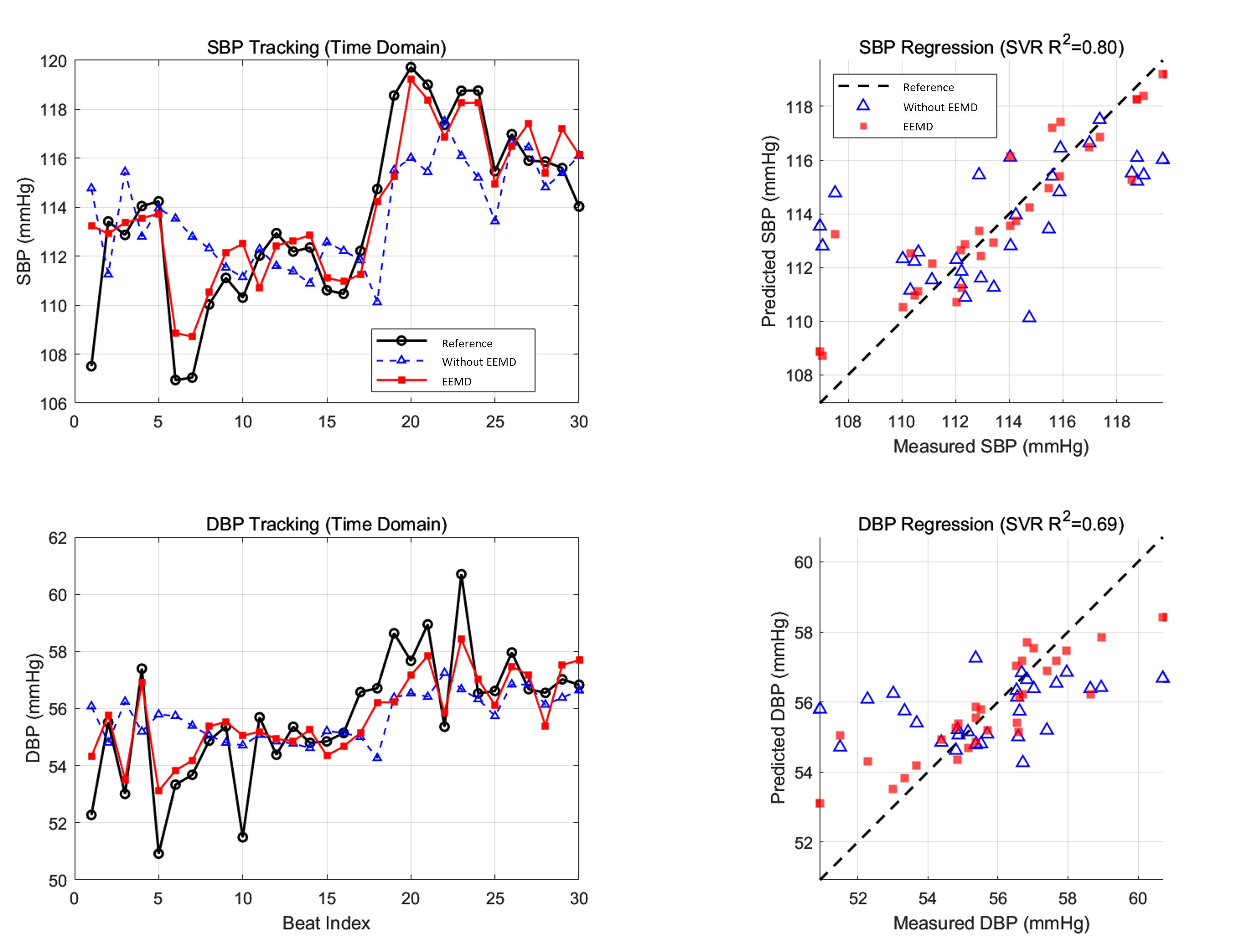}
    \caption{\textbf{EEMD Effectiveness.} (Right) Regression results confirm improved precision (tighter clustering).}
    \label{fig:tracking}
\end{figure}

\subsection{Ablation Study: Impact of Viscoelastic Compensation}
\label{subsec:ablation}

To isolate the specific contribution of the proposed physics-informed EEMD features, we conducted an ablation study comparing the full model against a baseline model that utilizes only standard elastic features (PTT, HR, Amp).

\subsubsection{Quantitative Improvement}
Table \ref{tab:ablation} summarizes the performance metrics on the test set. The baseline model yields significantly higher error margins. In contrast, the proposed method reduces the SBP RMSE by approximately 20\%. This substantial improvement confirms that high-frequency IMFs successfully capture information orthogonal to the standard PTT, effectively calibrating the non-linear relationship between vessel wall movement and pressure changes.

\begin{table}[t]
    \caption{Ablation Study: Performance Comparison}
    \begin{center}
    \begin{tabular}{lcccc}
    \toprule
    \multirow{2}{*}{\textbf{Model Configuration}} & \multicolumn{2}{c}{\textbf{SBP}} & \multicolumn{2}{c}{\textbf{DBP}} \\
    \cmidrule(lr){2-3} \cmidrule(lr){4-5}
     & \textbf{RMSE} & \textbf{MAE} & \textbf{RMSE} & \textbf{MAE} \\
    \midrule
    Baseline (Without EEMD) & 6.58 & 4.65 & 3.94 & 2.60 \\ 
    \textbf{Proposed (With EEMD)} & \textbf{5.22} & \textbf{3.70} & \textbf{3.65} & \textbf{2.56} \\
    \bottomrule
    \end{tabular}
    \label{tab:ablation}
    \end{center}
\end{table}

\subsubsection{Dynamic Tracking Analysis}
The limitations of the baseline model become most apparent during rapid hemodynamic fluctuations. Fig. \ref{fig:tracking} presents a beat-to-beat tracking case study.

As observed in the time-domain plots, the baseline model exhibits a distinct "damping effect" or hysteresis. Specifically, during the sharp SBP elevation (e.g., beats 15-20), the baseline fails to reach the peak pressure, resulting in significant under-estimation. This behavior is consistent with the physical nature of viscoelastic materials, which resist rapid deformation.

However, the proposed model, equipped with the EEMD velocity metric, accurately tracks these transient spikes. By compensating for the viscous damping term ($\eta \cdot \frac{d\epsilon}{dt}$), our method "accelerates" the model's response, allowing it to tightly follow the ground truth even when BP fluctuates by $>15$ mmHg within a few beats. This capability is critical for detecting acute hypertensive events in real-time monitoring.

\section{Conclusion}
\label{sec:conclusion}

This work challenges the oversimplified elastic assumption in cuffless BP monitoring by introducing a physics-informed Kelvin-Voigt viscoelastic framework. By leveraging EEMD to isolate viscous damping components ($\eta \cdot \dot{\epsilon}$), our model achieves AAMI-compliant accuracy (RMSE: 5.22/3.65 mmHg) on a challenging, high-hypertension MIMIC-II subset. Crucially, the method significantly mitigates hysteresis during rapid fluctuations. We conclude that viscoelastic compensation is not merely an algorithmic refinement but a physiological necessity for robust, medical-grade monitoring. Future work will extend this framework to ambulatory settings with multi-modal sensing.

\end{document}